\documentclass[aps,prb,twocolumn,floatfix]{revtex4-2}
\usepackage{graphicx}
\usepackage[colorlinks=true,linkcolor=blue,citecolor=blue,urlcolor=blue]{hyperref}
\usepackage{siunitx}
\DeclareSIUnit\rydberg{Ry}
\DeclareSIUnit\atomicunit{a.u.}
\DeclareSIUnit\bohr{\text{\ensuremath{a_0}}}
\usepackage[all]{hypcap}
\usepackage{float}
\usepackage[english]{babel}

\usepackage{amsmath}
\usepackage{amssymb}
\usepackage{xcolor}

\newcommand{\VEC}[1]{\mathbf{#1}}

\begin{document}

\title{Frustrated magnetism in Mn films on Ag(111) surface: from chiral in-plane N\'eel state to row-wise antiferromagnetism }

\author{Selcuk Sözeri$^{1,2}$, Nihad Abuawwad$^{2,1}$, Amal Aldarawsheh$^{2,1}$,  Samir Lounis$^{2,1}$}
\address{$^1$ Faculty of Physics, University of Duisburg-Essen and CENIDE, 47053 Duisburg, Germany}
\address{$^2$ Peter Gr\"unberg Institut,  Forschungszentrum J\"ulich \& JARA, 52425 J\"ulich, Germany}
\date{\today}%

\begin{abstract}
We conduct a comprehensive density functional theory (DFT) study to explore the intricate magnetic properties of frustrated Mn monolayer on the Ag(111) surface. Spin-polarized scanning tunneling microscopy demonstrates that a N\'eel magnetic state characterizes such an interface, which contradicts systematic ab-initio predictions made in the last two decades indicating that the ground state is collinear row-wise antiferromagnetic (RW-AFM) state. Here, we employ the all-electron full-potential Korringa-Kohn-Rostoker Green function (KKR) method and find that the ground state is a chiral magnetic N\'eel  state, with magnetic moments rotating in the surface plane following a unique sense of rotation, as dictated by the underlying in-plane magnetic anisotropy and Dzyaloshinskii-Moriya interaction. Once allowing disordered magnetic states, as described within the disordered local moment (DLM) approach, we reveal the possibility of stabilization of a RW-AFM state. We conjecture that at low temperatures, the chiral N\'eel state prevails, while at higher temperatures, the magnetic exchange interactions are modified by magnetic disorder, which can then induce a transition towards a RW-AFM state. Our work addresses a long term experimental-theoretical controversy and provides significant insights into the magnetic interactions and stability of Mn films on noble metal substrates, contributing to the broader understanding of the different magnetic facets of frustrated magnetism in thin films.
\end{abstract}

\maketitle

\section{Introduction}

Frustrated magnets, where competing interactions between neighboring spins cannot be simultaneously satisfied, give rise to complex and intriguing phenomena. This frustration creates a highly degenerate ground state and lies at the heart of noncollinear antiferromagnetism, giving rise to exciting properties such as Hall effects \cite{Nakatsuji2016LargeAH}, torques \cite{Yutaro}, and topological spin-textures \cite{AMAL_Multimeronic_2023,Aldarawsheh2023model,PhysRevB.108.094409}.
A classic example is a 2D hexagonal lattice of antiferromagnetic atoms, where the topological constraints make it impossible to align all neighboring spins in an antiparallel arrangement. In certain conditions, frustration can lead to the formation of spin liquid states, exhibiting remarkable collective behaviors, in which spins remain highly correlated but continue to fluctuate, even at absolute zero \cite{Balents:2010wrb}.  The study of these frustrated spin structures is not only fundamental to understanding exotic magnetic states but also critical for advancing spintronics applications \cite{Rimmler}.

\begin{figure*}[tb]
\includegraphics[width=\linewidth]{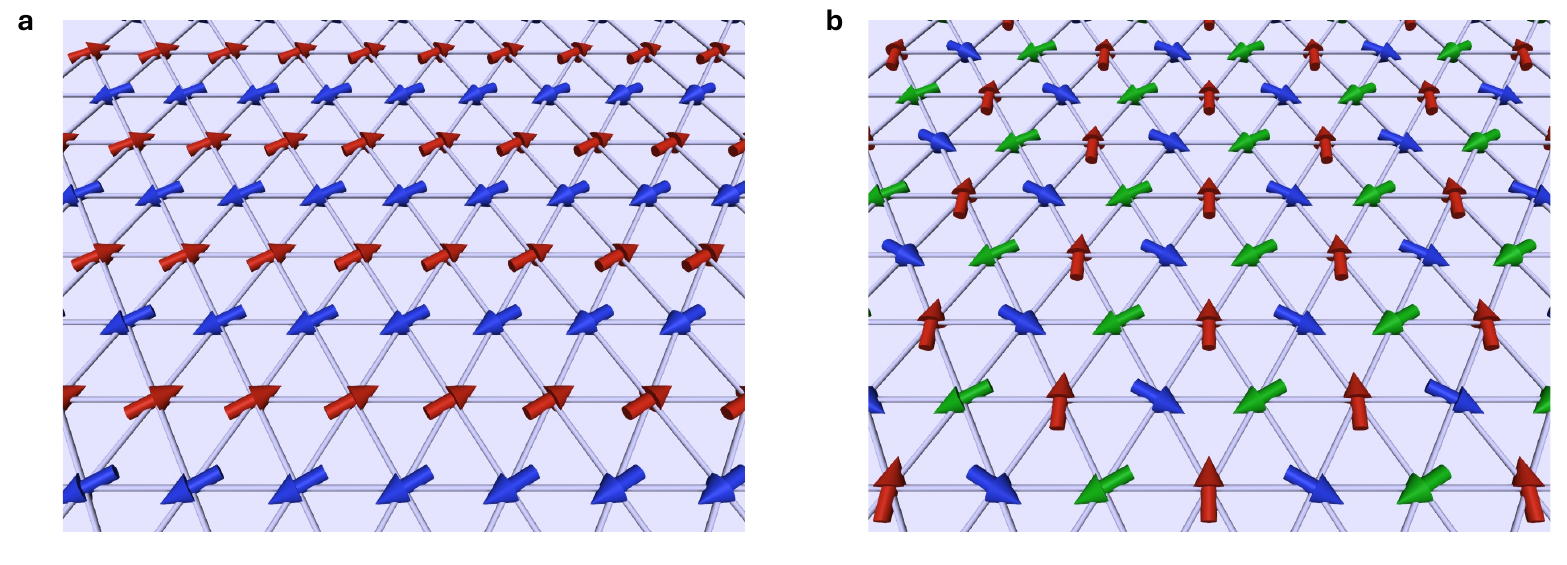}
\caption{{ a) Illustration of a row-wise antiferromagnetic (RW-AFM) state on a triangular lattice, where adjacent rows exhibit alternating spin orientations.
b) Illustration of a Néel state on a triangular lattice, characterized by a spin arrangement where neighboring spins are aligned in a non-collinear configuration with a rotation angle of 120$^\circ$, minimizing frustration in the triangular geometry.} 
   }
\label{FIG}
\end{figure*}

The exploration of magnetic frustration in the limit of two-dimensional materials has been pursued since decades. A two-dimensional hexagonal (or triangular) arrangement of antiferromagnetic atoms is typically achieved by depositing a single monolayer (ML) of magnetic elements, expected to be antiferromagnetic such as manganese (Mn), onto hexagonal surfaces initially non-magnetic, e.g. (111) surfaces of face-centered cubic (fcc) crystals. For instance, a Mn ML on Ag(111) surface has been the subject of extensive study over the years, encompassing both theoretical and experimental approaches. Intuitively and due to magnetic frustration induced by nearest neighboring (n.n.) antiferromagnetic (AFM) interactions, a 120$^\circ$ N\'eel state is anticipated as the ground state, see the comparison illustrated in Fig.~\ref{FIG} between a N\'eel state and a row-wise antiferromagnetic (RW-AFM) configuration. In the former, the n.n. magnetic moments are rotated by an angle of 120$^\circ$. However, early first-principles  investigations~\cite{PKurz,heinze} based on the Full-potential Linearized Augmented Plane Wave (FLAPW) method predicted a different ground state for Mn/Ag(111): a row-wise antiferromagnetic (RW-AFM) state (Fig.~\ref{FIG}a). The latter was consistently supported by subsequent studies employing various ab-initio methods based on density functional theory (DFT). Notable examples include the use of pseudopotentials~\cite{MALONDABOUNGOU2019e00368}, tight-binding linear muffin-tin orbitals~\cite{PhysRevB.61.15277} and the all-electron Korringa-Kohn-Rostoker (KKR) Green function formalism in the atomic sphere approximation~\cite{PhysRevB.83.054435}. 

Despite the consistency of these theoretical findings, state-of-the-art spin-polarized scanning tunneling spectroscopy measurements undoubtly indicate that a Mn ML grown on Ag(111) hosts a Néel magnetic state~\cite{PhysRevLett.101.267205}. This longstanding discrepancy between theoretical predictions and experimental observations, persisting for several decades, presents an intriguing puzzle and underscores the complexities of modeling such systems. More recently, the N\'eel state of a Mn monolayer has been observed on a different substrate, the Ir(111) surface~\cite{doi:10.1021/acsnano.3c11459}, which as predicted theoretically can host frustrated topological spin-textures~\cite{AMAL_Multimeronic_2023}.

In this paper, we aim to address the persistent issue in describing the magnetic ground state of a Mn ML on Ag(111). We compare our findings to previous studies and discuss the decisive mechanisms for the emergence of the N\'eel state. While we recover the measured magnetic state utilizing the all-electron full-potential KKR method, we demonstrate the chiral nature of the N\'eel state and explore the impact of magnetic disorder within the disordered local moment (DLM) approach. The latter enables to mimic the impact of temperature, and shows that the RW-AFM state can be recovered if disorder is incorporated.

\section{Methods}

\subsection{First-principles calculations}

For atomic relaxations, we employed the projected augmented wave (PAW) pseudopotential \cite{ps} as implemented in the ab initio simulation package Quantum Espresso (QE) \cite{qe}, which is firmly rooted in DFT. 
We assumed a k-mesh resolution of 20×20×1, while the plane-wave and energy cutoffs are considered to be 110 Ry and the convergence criterion for the total energy is set to 0.01 $\mu$Ry. During the relaxation process, the atomic positions were optimized to ensure that the residual forces on the atoms were less than \(1 \, \text{mRy} \, a_0^{-1}\).

Once the geometry is optimized, we explore the magnetic properties adopting the DFT-based all-electron full-potential KKR method, utilizing the JuKKR code~\cite{russmann2022judftteam} where spin-orbit coupling is incorportated self-consistently. The method enables the exploration of non-collinear structures as well as the inclusion of the coherent potential approximation (CPA), which permits the investigation of magnetic disorder in the so-called DLM approach~\cite{BLGyorffy_1985,PhysRevLett.69.371,PhysRevB.74.144411}. DLM describes how the electronic structure responds to thermal fluctuations of the local moments. The Mn layer is then assumed to be an alloy Mn$_{p}$Mn$_{1-p}$ made of magnetic moments pointing either in one or the opposite direction, with $p=50\%$ being the paramagnetic state expected at high temperatures. We treated the exchange-correlation interactions with the local density approximation (LDA), as formulated by Vosko, Wilk, and Nussair \cite{Vosko1980}.

Similarly to the pseudo-potential method, we employed a k-points grid mesh of 20×20×1 for the self-consistent calculations.  The angular momentum expansion of the Green function was truncated at lmax=3 and the energy integration were performed along the complex plane including a Fermi–Dirac smearing of 502.78 K. To extract the exchange interaction parameters such as the Heisenberg exchange and the Dzyaloshinskii-Moriya interaction (DMI)~\cite{DZYALOSHINSKY1958241,PhysRev.120.91}, we used the infinitesimal rotation method~\cite{LIECHTENSTEIN198765,PhysRevB.79.045209}  with a finer k-points grid mesh of 200×200×1. 

\subsection{Magnetic interactions and atomistic spin dynamics}
The magnetic interactions obtained from the first-principles calculations are used to parameterize the following classical extended Heisenberg Hamiltonian :
\begin{eqnarray}\label{eq:spin_model}
\begin{aligned}
H &= - \sum_{i,j} \VEC{S}_i \cdot \mathcal{J}_{ij} \cdot \VEC{S}_j - \sum_i K_i (S_i)^2  \\
  &= - \sum_{i,j} J_{ij} \VEC{S}_i \cdot \VEC{S}_j - \sum_{i,j} \VEC{D}_{ij} \cdot (\VEC{S}_i \times \VEC{S}_j) - \sum_i K_i (S_i^z)^2\;.
\end{aligned}
\end{eqnarray}

Here $i$ and $j$ label different magnetic sites carrying a magnetic moment, whose orientation is defined by the unit vector $\mathbf{S}$, while $\mathcal{J}$ is the tensor of exchange interactions. The elements of the tensor  can be rearranged to form the isotropic Heisenberg exchange coupling $J$, the DMI $D$. The positive sign of the magnetic anisotropy energy $K$  indicates a preference for an out-of-plane orientation of the moments.

After extraction of the magnetic interactions, a Fourier-transformation is performed: $\mathcal{J}_{ij}(\VEC{q}) = \sum_j \mathcal{J}_{0j}e^{-i\VEC{q}\cdot\VEC{R_{0j}}}$, where $\VEC{R}_{0j}$ is a vector connecting unit cells atom 0 and $j$. By evaluating the associated eigenvalues when obtains the dispersion curves of spin spirals characterized by the wave vector $\VEC{q}$. 

Furthermore, atomistic spin dynamic simulations using the Landau-Lifshitz-equation (LLG) as implemented in the Spirit code \cite{Mueller2019a}\footnote{The Spirit code can be found at \url{https://juspin.de}} are performed in order to explore the existence and stability of various trivial and complex magnetic states.  Once the magnetic ground state was found with the atomistic spin model, the stability of the spin-texture has been once more studied via DFT.

\section{Results and discussion}

\subsection{Atomic relaxations}  

We consider two scenarios for the stacking of the Mn ML on Ag(111), either fcc or hcp.  The Mn ML relaxes towards the Ag interface by 
$2.48\%$ (fcc) and $2.29\%$ with respect to the Ag ideal interlayer distance. We find that the fcc stacking is energetically more favorable than the hcp one with an energy difference of 6.83 meV. Consequently, and since the magnetic exchange interactions as well as the magnetic anisotropy energy (MAE) are found similar for both stacking cases, the subsequent analyses focuses exclusively on the fcc configuration.

\subsection{Heisenberg exchange interactions}
\begin{figure*}[!htb]
\includegraphics[width=\linewidth]{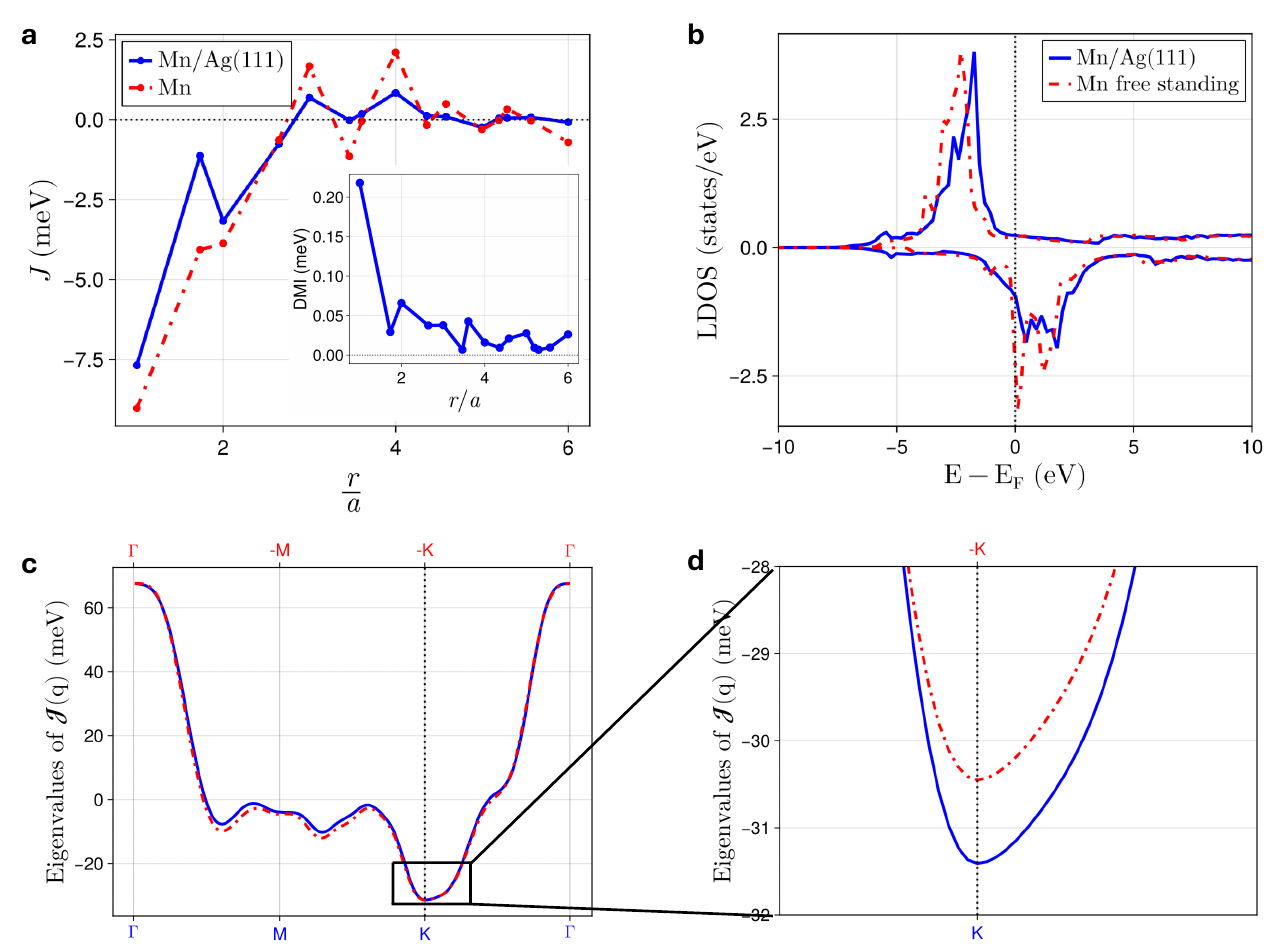}
\caption{(a) The Heisenberg exchange interactions as a function of distance for Mn/Ag(111) (blue) and the unsupported Monolayer Mn (red) while the inset shows the DM interactions. 
(b) Spin-resolved local density of states (LDOS) of Mn/Ag(111) (blue) and Mn (red). 
 (c) Eigenvalues of the Fourier-transformed tensor of exchange interactions high symmetry paths for Mn/Ag(111). The lowest values indicate the ground state. The DMI-induced splitted blue and red curves in (c) correspond to spin-spirals along opposite high-symmetry directions. }
   
\label{FIG1NEW}
\end{figure*}
We now turn our attention to the analysis of  the magnetic exchange interactions as function of interatomic distances, illustrated in Fig.~\ref{FIG1NEW}a. 
Specifically, we investigate two scenarios: first, a free-standing hexagonal Mn monolayer with the imposed Ag lattice parameter (as done previously in Refs.~\cite{PKurz,heinze}), and second, a configuration in which the Mn layer is  supported by the Ag(111) substrate assuming an fcc stacking. Fig.~\ref{FIG1NEW}a clearly indicates the presence of significant antiferromagnetic Heisenberg exchange interactions up to forth nearest neighbors. At larger distances, the interactions oscillate as  expected for Ruderman–Kittel–Kasuya–Yosida (RKKY) interactions in metallic materials. The figure unequivocally demonstrates that the hybridization with the substrate affect the Heisenberg exchange interactions.

Notably, the exchange interactions involving the first, second, and third nearest neighbors are markedly reduced when manganese is interfaced with silver. As it will be shown later, this reduction is a critical factor driving the transition of the system's ground state from one magnetic state to a totally different one.

Armed with the magnetic interactions, we can explore the magnetic ground state dictated by the tensor of exchange interactions. This can be studied  by analysing the eigenvalues of the Fourier-transformed exchange interactions illustrated in Fig.~\ref{FIG1NEW}c. The minimum is located at the K-point, which corresponds to the N\'eel magnetic state, which is in accordance with the SP-STM measurements conducted by Gao et al.~\cite{PhysRevLett.101.267205}. Although being the ground state, one identifies a  surrounding flat energy curve, which expresses the possibility of finding energetically close metastable spin-spiraling states of antiferromagnetic nature. Note that in Fig.~\ref{FIG1NEW}c two dispersions curves are shown, which corresponds to spirals propogating along opposite high-symmetry directions of the Brillouin zone. The breaking of symmetry is induced by the presence of DMI vectors as elaborated in the next subsection. The energy difference with respect to the RW-AFM state amounts to 27.5 meV. 

\begin{figure*}[tb]
\includegraphics[width=\linewidth]{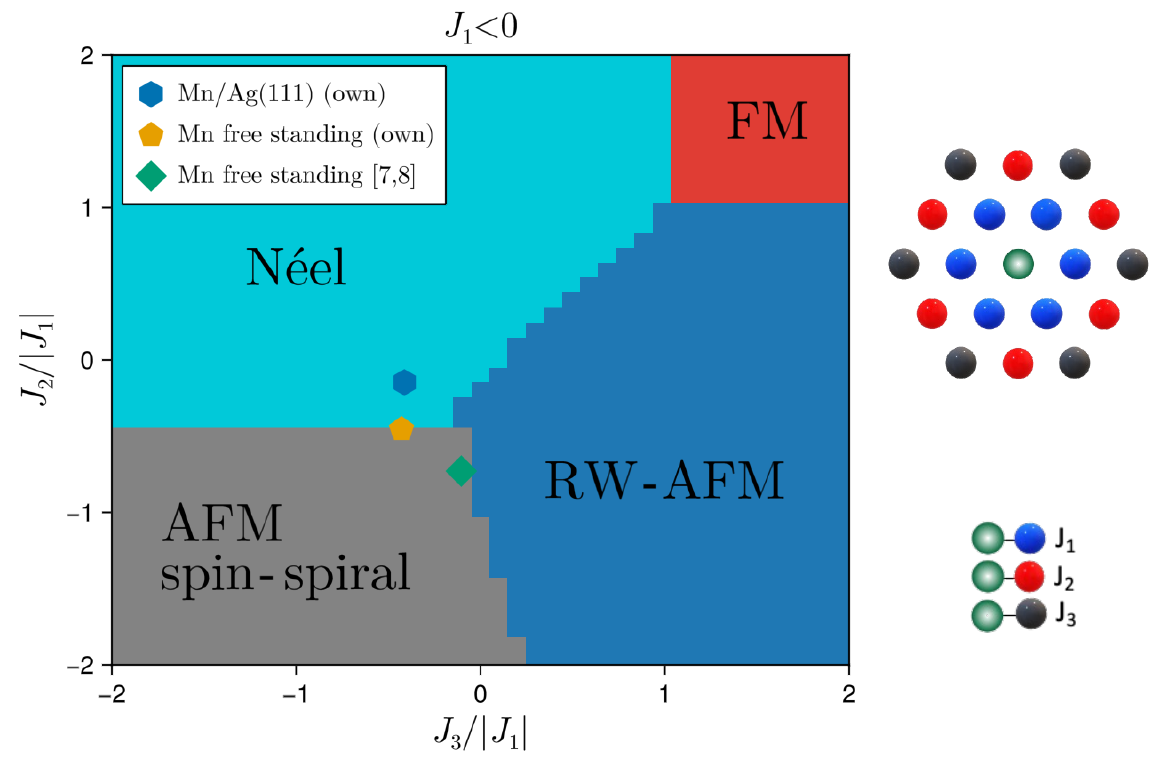}
\caption{\label{figure_1} 
Phase diagram for the Heisenberg model on a triangular lattice without spin–orbit-induced interactions, considering magnetic exchange interactions up to the third nearest neighbors illustrated on the right-hand side for a two-dimensional hexagonal lattice. $J_1$ (assumed antiferromagnetic), $J_2$ and $J_3$ are the interactions between the central atom and the first (blue), second (red) and third (black) n.n. atoms, respectively. Four phases emerge: ferromagnetic (FM), RW-AFM, N\'eel, and AFM spin-spirals. Our simulations locate the ground state for Mn/Ag(111) in the N\'eel sector while the Mn free standing case hosts AFM spin-spirals. The result of previous simulations~\cite{PKurz,heinze} when limited to the three n.n. interactions is also indicated.
   }
\label{FIG2}
\end{figure*}

At this stage, it is instructive to examine the phase diagram of an antiferromagnetic hexagonal monolayer, considering three n.n.  interactions, in the same spirit as presented in Refs.~\cite{PKurz,heinze}. Assuming an AFM first n.n. interaction $J_1$, we obtain the phase diagram plotted in Fig.~\ref{FIG2}. One notices that for sufficiently large ferromagnetic  second ($J_2$) and third ($J_3$) n.n. interactions, the ground state can be ferromagnetic despite the AFM $J_1$.  Starting from that magnetic phase and reducing $J_2$ while keeping $J_3$ ferromagnetic give rise to the RW-AFM ground state observed experimentally  with SP-STM for an fcc-stacked Mn/Re(0001)~\cite{Paper_Mn_RE_0001}. Reversing the sign of $J_3$   triggers a transition to an antiferromagnetic (AFM) spin-spiral state. By further weakening $J_2$  starting from this state, the N\'eel state becomes stabilized. Based on the magnetic interactions derived from our simulations, and considering up to third n.n.  interactions, the N\'eel state emerges as the ground state for the Mn layer on Ag(111). Notably, a similar state was predicted for a Cr layer interfaced with a PdFe bilayer grown on an Ir(111) surface \cite{Aldarawsheh:963986}.  

Removing the substrate enhances the relative strength of $J_2$  compared to $J_3$, thereby stabilizing the AFM spin-spiral phase near the boundary with the N\'eel phase. This behavior is directly linked to changes in the electronic structure induced by deposition on the Ag(111) surface, as illustrated in Fig.~\ref{FIG1NEW}b. The exchange splitting increases when the substrate is removed, which leads to an enhancement of the magnetic moment and effective n.n. coupling, while reducing relatively the interaction between the second n.n. atoms. Utilizing the previously published simulations~\cite{PKurz,heinze}, the ground state shifts from the RW-AFM configuration to the spin-spiraling phase (close to the boundary with the RW-AFM realm) when limiting the interactions to the third n.n. interactions (Fig.~\ref{FIG2}).

\begin{figure*}[!htb]
\includegraphics[width=\linewidth]{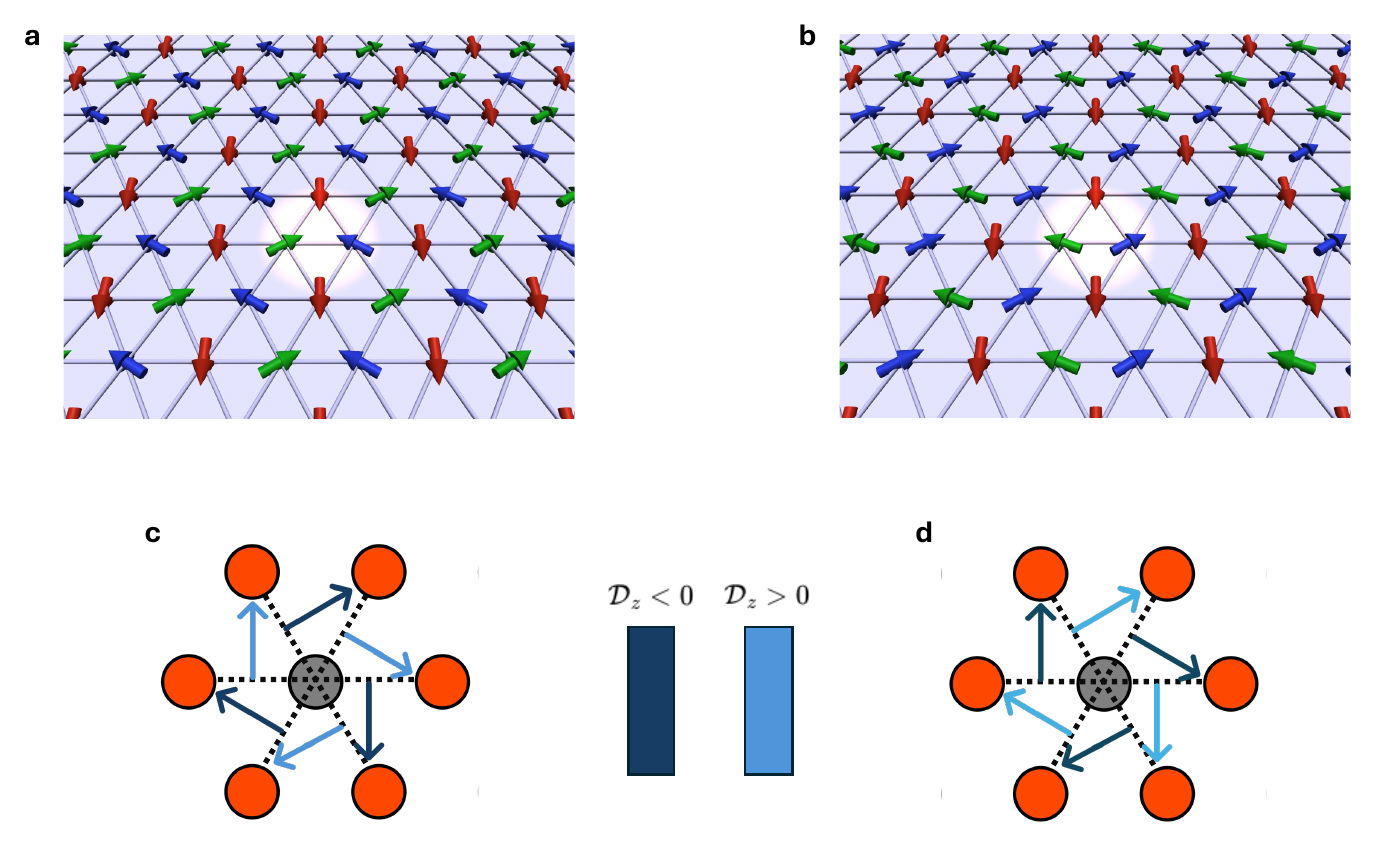}
\caption{In-plane N\'eel states of different chiral nature. (a) The ground states dictated by the DMI vectors as obtained from our DFT simulations. The n.n. DMI vectors rotate clockwise as shown in (c). (b) The chiral nature of the N\'eel state changes when flipping the sign of the z-component of the DMI vector (shown in (d)). The latter component enforces a rotation of the moments within the surface plane, with its sign determining the direction of the rotation.  The highlighted trimers in (a) and (b) clearly shows the distinct nature of their respective chiral states. The red magnetic moment is fixed, while the other ones have their directions switched. 
   \label{FIG3}
   }
\end{figure*}

\subsection{Impact of spin-orbit coupling \& spin chirality}

The magnetic anisotropy energy (MAE) is found to be favoring an in-plane orientation of the moments for both stacking cases. The MAE reaches a value of -0.13 meV. Besides the MAE, SOC of the Ag surface gives rise to substantial DMI, as plotted in the inset of Fig.~\ref{FIG1NEW}a.  The DMI vectors mediating chiral interactions among nearest neighbors rotate in a clockwise fashion (Fig.~\ref{FIG3}c) with an amplitude of 0.22 meV. This gives rise to the chiral N\'eel state plotted in Fig.~\ref{FIG3}a, which is more stable than its chiral counterpart illustrated in Fig.~\ref{FIG3}b by an energy of 0.95 meV. The latter is obtained by flipping the sign of the z-component of the DMI vectors (Fig.~\ref{FIG3}d), which switches the direction of two moments belonging to the trimers highlighted in Figs.~\ref{FIG3}(a,b). Dark (light) blue DMI vectors in (c) and (d) correspond to negative (positive) DMI vector's z-component.  
The impact of the chiral interaction is clearly seen in Fig.~\ref{FIG1NEW}c, 
where the degenearcy is lifted for the eigenvalues of the Fourier-transformed exchange interactions are plotted as a function of the wave vector q along the two opposite high-symmetry q-paths:  $\Gamma$→M→K→$\Gamma$ versus $\Gamma$→-M→-K→$\Gamma$. 

The energy difference between the two chiral Néel states is consistently reproduced, both using the spirit code and the eigenvalues of $\mathcal{J}$(q) plot. This consistency across different methods underscores the robustness of the energy difference calculatios. Thus, according to our simulations, Mn/Ag(111) hosts a chiral N\'eel state as a magnetic ground state, with moments rotating in-plane.

\begin{table*}[!ht]
\centering
\caption{\label{table:DLM} Spin magnetic moments $m$, magnetic exchange interactions $J$, magnetic anisotropy energies ($K$) and Dzyaloshinskii-Moriya interaction ($D$) of the Mn monolayer fcc stacked on Ag(111) surface. The energy differences between the RW-AFM and the two chiral N\'eel states 1 (Fig.~\ref{FIG3}a) and 2 (Fig.~\ref{FIG3}b) are listed, with negative values indicating that the N\'eel state is lower energetically.} 
\begin{ruledtabular}
\bgroup
\def\arraystretch{1.5}
\begin{tabular*}{\textwidth}{c @{\extracolsep{\fill}} rrrrrr}
\multicolumn{7}{c}{Mn/Ag(111) for different concentration} \\ \cline{2-7} 
 & 100\% & 90\%& 80\% & 70\% &60\% & 50\% \\
\hline
$m$ ($\mu_B$) & 4.07 & 4.06 & 4.04 & 4.03 & 4.02 & 4.03 \\
$K$ (meV) & -0.13 & -0.05 & -0.02 & 0.00 & -0.01 & 0.01 \\
${J}_1$ (meV) & -7.77 & -7.56 & -7.38 & -7.36 & -7.53 & -7.88 \\
${J}_2$ (meV) & -1.34 & -1.95 & -2.44 & -2.75& -2.87 & -2.82 \\
${J}_3$ (meV) & -3.36 & -1.61 & -0.63 & -0.03 & 0.30 & 0.42 \\
$|\mathbf{D}_1|$ (meV) & 0.25 & 0.17 & 0.15 & 0.14 & 0.16 & 0.19 \\
$|\mathbf{D}_2|$ (meV) & 0.06 & 0.01 & 0.00 & 0.02 & 0.05 & 0.07 \\
$|\mathbf{D}_3|$ (meV) & 0.06 & 0.04 & 0.02 & 0.01 & 0.00 & 0.01 \\
$\Delta E_\text{N\'eel1 - RW-AFM}$ (meV) &  {-27.50} &-8.79  &4.24  &12.61 &{16.69}  & {17.22}  \\
$\Delta E_\text{N\'eel2 - RW-AFM}$ (meV) &  -26.55 &-9.40  &3.95  &{12.49}  &{16.69}  & 17.35  \\
\end{tabular*}
\egroup
\end{ruledtabular}
\end{table*}

\subsection{Disordered local moment approximation}
Next, we examine the influence of temperature on the exchange interaction within the DLM framework~\cite{BLGyorffy_1985,PhysRevLett.69.371,PhysRevB.74.144411,Tapia2021}. As the temperature increases, spin fluctuations are induced, which progressively disrupt the long-range magnetic order and ultimately diminish the overall spin polarization of the system's electronic structure. As aforementioned in the method section, magnetic disorder is incorporated in terms of concentration of magnetic moments pointing in one or the opposite direction on each Mn site.

\begin{figure*}[!htb]
\includegraphics[width=1.\linewidth]{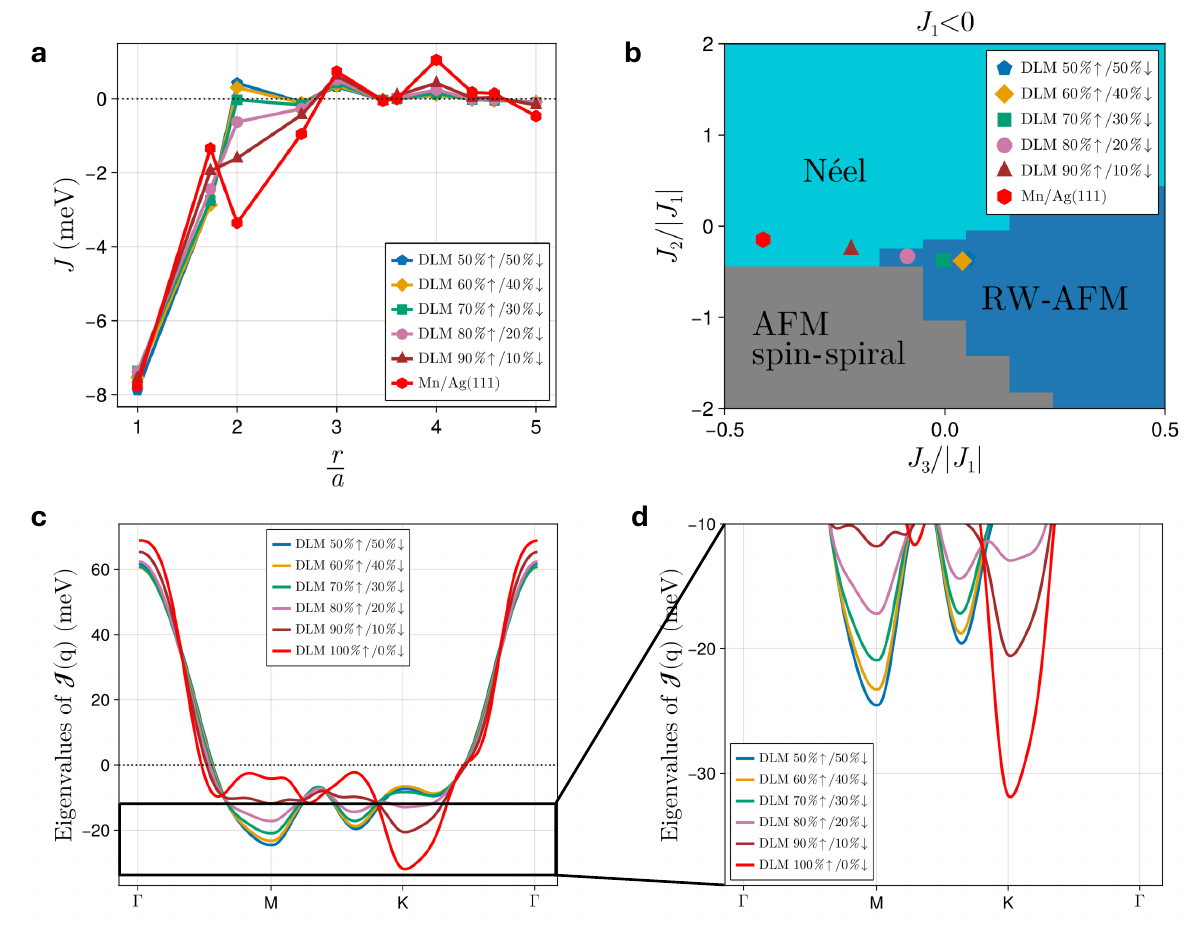}
\caption{ (a) The Heisenberg exchange interactions as a function of distance for different DLM concentrations. 
(b) Phase diagram including the three n.n.  interactions focusing on the impact of magnetic disorder on the ground state. \textbf{c)} Eigenvalues of the Fourier-transformed exchange interactions as a function of q along the high-symmetry paths  $\Gamma$→M→K→$\Gamma$. The lowest values indicate the ground state.}
   \label{FIG4}
\end{figure*}  

In Fig.~\ref{FIG4}a, we illustrate this effect by visualizing different concentrations, ranging from 50$\%$ (representing the paramagnetic state) to 100$\%$ (representing the fully ordered state), in increments of 10$\%$. 50$\%$  corresponds to the case that on each site the magnetic moment points either upwards or downwards with equal concentration.

 The results demonstrate  how increasing magnetic disorder, which could be related to an increase in temperature,  affects the exchange interactions. The values of the interactions up to the third nearest neighbors, including the DMI as well as the MAE are reported in Table \ref{table:DLM}. The trends are clear. Interestingly, while the spin moments are stable, the MAE gets dramatically reduced as soon as a bit of magnetic disorder is introduced. Thus the in-plane orientation of the moments is not maintained. 
 In contrast to the n.n. Heisenberg and DMIs, which are more or less constant independently from the magnetic disorder, the second and third n.n.  interactions experience a dramatic impact. For instance, $J_2$ increases in magnitude, while $J_3$  decreases and changes even sign below 70\% magnetic disorder. This leads to a shift of the ground state from the N\'eel state towards a RW-AFM state as illustrated in the phase diagram plotted in Fig.~\ref{FIG4}b. We conjecture that by increasing temperature, the magnetic interactions can be affected, more importantly those mediating the coupling among the 3rd n.n. atoms, which would lead to a phase transition towards a collinear RW-AFM state. 

 We note that the sense of rotation of the magnetic moments can change as function of magnetic disorder. The  N\'eel ground state is identical for the ordered and paramagnetic cases (Fig.~\ref{FIG3}a), but switches to its chiral image for the remaining DLM states (Fig.~\ref{FIG3}b).

The results obtained with the simple Heisenberg model leading to the aforementioned phase diagram are confirmed when including magnetic interactions beyond nearest neighbors.
Fig.~\ref{FIG4}c presents the eigenvalues of the Fourier-transformed exchange interactions as a function of the wave vector q, considering different concentrations within the DLM model. The energy differences between the two  N\'eel states and RW-AFM are listed in Tab.~\ref{table:DLM}.

 For magnetic disorder between 50$\%$  and 80$\%$, the global minimum of the eigenvalues is located at the M point in the Brillouin zone, which corresponds to the RW-AFM ground state, suggesting that at these concentrations, the RW-AFM configuration is energetically preferred. As the concentration increases to 90$\%$  and above, a notable transition occurs. The global minimum shifts from the M to the K point, which corresponds to the Néel state. This transition signifies a change in the preferred magnetic ground state when tuning magnetic disorder.

\section{Conclusions}

We performed extensive ab-initio simulations to address the magnetic ground state of Mn monolayer on Ag(111) surface. This AFM material was previously predicted from ab-initio to host a RW-AFM ground state, which was contradicted by SP-STM measurements that demonstrate a N\'eel magnetic state. Our work addresses a long-standing puzzle in magnetic thin films.

Utilizing the all-electron KKR Green function method, we find the Mn monolayer being stacked in either an fcc or hcp sites to be characterized by a N\'eel magnetic state in total accordance with experimental measurements. Furthermore, the ground state is found chiral  in-plane as dictated by the underlying DMI induced by the presence of the heavy Ag surface atoms and the in-plane magnetic anisotropy energy. We analysed our data and explain them in terms of the extracted magnetic interactions and electronic structure. Since small changes in the magnetic interactions can switch the ground state from the N\'eel to the RW-AFM state, this enables an intuitive explanation of the so-far reported ab-initio simulations.

Subsequently, we extended our study to explore the impact of magnetic disorder utilizing the DLM approach, which would mimic the impact of temperature on the electronic structure and therefore on the associated magnetic interactions. We learned that magnetic disorder affects mainly the MAE, by reducing it dramatically, while changing the sign of the 3rd n.n. Heisenberg exchange interaction. The latter aspect is responsible for stabilizing the RW-AFM state when magnetic disorder kicks in. We conjecture that at low temperature, as measured with SP-STM, the magnetic ground state is a chiral in-plane N\'eel state, while at higher temperature a phase transition is expected which leads to the formation of a RW-AFM state.

\begin{acknowledgments}
We acknowledge fruitful discussions with Hangyu Zhou, Manuel dos Santos Dias, Jens Wiebe, Lucas Schneider and Gustav Bihlmayer. 
Funding has been provided by the Priority Programmes SPP 2137 “Skyrmionics” (Projects LO 1659/8-1) of the Deutsche Forschungsgemeinschaft (DFG). 
Computations were performed with computing resources granted by RWTH Aachen University under project p0020362.
\end{acknowledgments}

\bibliography{references.bib}

\end{document}